\documentclass[twocolumn,showpacs,preprintnumbers,amsmath,amssymb]{revtex4}
\usepackage{graphicx}
\usepackage{dcolumn}
\usepackage{bm}
\usepackage{epsfig}

\begin{document}


\title{A novel method of data analysis for hadronic physics}

\author{ C.N.~Papanicolas$^{1,2}$\footnote{corresponding author,
e-mail address: cnp@cyi.ac.cy}, E.~Stiliaris$^1$}

\affiliation{$^1$Department of Physics, University of Athens, Athens, Greece}
\affiliation{$^2$ The Cyprus Institute, Nicosia, Cyprus}

\date{\today}

\begin{abstract}
A novel method for extracting physical parameters from
experimental and simulation data is presented. The method is based
on statistical concepts and it relies on Monte Carlo simulation
techniques. It identifies and determines with maximal precision
parameters that are sensitive to the data. The method has been
extensively studied and it is shown to produce unbiased
results. It is applicable to a wide range of scientific
and engineering problems. It has been successfully applied in the
analysis of experimental data in hadronic physics and of lattice QCD
correlators.
\end{abstract}


\maketitle

A  principal task in experimental and computational physics
concerns the determination of the parameters of a theory (model)
from experimental or simulation data. Examples are abundant: the
determination of the parameters of the Standard Model in Particle
Physics, the multipole amplitudes in Nucleon Resonance excitation
in hadronic physics, the determination of the spectrum from correlators
in lattice QCD calculations, the parameters of acoustic resonances
in Cosmology, to mention a few.

The data from which information on parameters of the theory is to be extracted are
characterized by statistical uncertainties and systematic errors,
are typically of limited dynamic range and sensitive only to a few
of the model parameters, rendering this task  difficult and often
intractable. Identifying which parameters of the model
can be determined from the available data is often difficult to
prejudge and their extraction without bias is often
impossible. Particularly hard is the determination of the systematic and model
uncertainties that ought to be assigned to the extracted values of
the parameters.

We address this problem via a method,
that we will refer to as the Athens Model Independent
Analysis Scheme, "AMIAS", which is capable of extracting theory (model)
parameters and their uncertainties from a set of data in a
rigorous, precise, and unbiased way. The methodology is first
presented and then subsequently
applied to two problems in hadronic physics, which
are used as demonstration cases. The two cases concern
A) the extraction of the mass spectrum of hadrons from Euclidean time correlators
in lattice QCD simulations and
B) the extraction of the multipole excitation amplitudes for the Nucleon
resonances and in particular that of the first excited state of
the Nucleon, the $\Delta(1232)$ resonance.

The AMIAS method is applicable to problems in which the
parameters to be determined are linked in an explicit way to the
data through a theory or model. There is no requirement that
this set of parameters are orthogonal; they can be subjected to constraints,
e.g. by requiring that unitarity is satisfied.
The method requires that  a quantitative criterion
for the "goodness" of a solution is chosen and thus far we have employed
the ${\chi^2}$ criterion.

For a given theory any set of values for its parameters,
satisfying its symmetries and constraints, provides a solution having a finite
probability of representing reality. This probability can be quantified through a comparison to the data being analyzed. Based on these concepts AMIAS can be formulated as follows:
\begin{description}
\item
  A set of parameters $A_1, A_2,...,A_N \equiv \{A_{\nu}\}$
  which completely and explicitly describes a process within a theory, can be determined from a
  data set $\{\textit{V}_{k} \pm \varepsilon_{k}\}$,
  produced by this process, by noting that
  any arbitrary set of values $\{a_{\nu}\}^{j}$ for these parameters constitutes a solution having a probability
  $P(j)$ of representing "reality" which is equal to:
  \begin{eqnarray}
     P(j)=\scriptsize{G}[\chi^{2}(j), \{a_{\nu}\}^{j}]
  \end{eqnarray}
  where
  $\scriptsize{G}$ is a function of the data and the parameters of the model and of $\chi^{2}$, where,
  \begin{eqnarray}
     \chi^{2}(j)=\sum_{k} \Big\{ \frac{(U^{j}_{k} -V_{k})}{\varepsilon_{k}} \Big\}^{2}
  \end{eqnarray}
  Thus $P(j)$ is a function of the $\chi^{2}$ resulting from the
  comparison to the  data $\{V_{k} \pm \varepsilon_{k}\}$
  of the predicted, by the theory, values $U^{j}_{k}$ by the $\{a_{\nu}\}^{j}$ solution.
\end{description}

In the case where we chose $\scriptsize{G} = e^{-\chi^2/2}$ the results obtained by AMIAS are
related to those obtained by  ${\chi^2}$ minimization methods and widely used and implemented in
a number of codes (e.g. MINUIT~\cite{minuit}). The results become identical if correlations among
the parameters of the theory are absent or ignored.

We call an ensemble $Z$ of such $a_{\nu}^{j}$ solutions \emph{Canonical
Ensemble of Solutions}, which has properties that depend only on
the experimental data set. Similarly a
\emph{Microcanonical Ensemble of Solutions} can be defined as the
collection of solutions which are characterized by
\begin{eqnarray}
\chi^{2}_{A} \leq \chi^{2}\leq \chi^{2}_{B}
\end{eqnarray}
where $\chi^{2}_{A}$ and $\chi^{2}_{B}$ define a sufficiently
narrow range in $\chi^{2}$ space.  A case of particular interest concerns the microcanonical ensemble
near the minimum $\chi^{2}$ value:
\begin{eqnarray}
\chi^{2}\leq \chi^{2}_{min}+C
\end{eqnarray}
where C is usually taken to be the constant equal to the effective degrees of freedom of the problem.

The extraction of the model parameters
$\{A_{\nu}\pm\delta A_{\nu} \}$   for a specific
set of data can be accomplished by employing the following
procedure:

\begin{itemize}
  \item  A canonical ensemble of solutions, is being constructed by randomly
  choosing values, $\{a_{\nu}\}^{j}$, for the  set of parameters $\{A_{\nu}\}$ of
  the theory within the allowed physical limits and by imposing the required
  constraints. Each set $\{a_{\nu}\}^{j}$ constitutes a point in the ensemble which is labeled by the $\chi^{2}$ value this solution generates when compared
  with the data. In the absence of constraints, any given model
  parameter $A_{\nu}$ will assume all allowed values with equal probability (equipartition
  postulate).
  \item To each point of the ensemble $\{a_{\nu}\}^{j}$
  a probability is assigned,  equal to
  $P(j)$. Following standard statistical concepts,
  the probability $\Pi(a_{\nu})$ of a parameter $A_{\nu}$  assuming a
  specific value $a_{\nu}$ in the range $(a_{\nu},a_{\nu}+\Delta a_{\nu})$  is equal to:
  \begin{eqnarray}
     \Pi(a_{\nu}) = \frac{\int\limits_{a_{\nu}}^{a_{\nu}+\Delta a_{\nu}}{\sum_j dA_{\nu}^j \ {P(j)}}} {\int\limits_{-\infty}^{+\infty}{\sum_j dA_{\nu}^j \ P(j)}}
  \end{eqnarray}
  This expression defines the Probability Distribution Function (PDF)  of any parameter of the theory for representing "reality".  It thus contains the maximum
  information that can be obtained from the given set of data. Having obtained the PDF, numerical results can be derived, usually moments of the distribution.
  The mean value is normally identified as the "solution" and the corresponding variance as its "uncertainty".
\end{itemize}

It is manifestly obvious that AMIAS has minimal assumptions and
that it introduces no methodological bias to the solution.
It determines the theory/model parameters that exhibit sensitivity to the
data yielding  PDFs that allow only a restricted range of values usually with a well defined maximum and a narrow width. If the data do not contain physical information to determine some of the parameters, then the resulting
PDFs are featureless. The underlying stochastic approach allows easy scalability
to a very large number of parameters even with limited number of data.

The method is computationally robust and stable; obviously it could not have been implemented
without the advent of powerful computers. We have developed algorithms that
can be implemented efficiently and produce results to realistic
and demanding problems within reasonable computation times. It is also apparent that the method is amenable to trivial computational parallelism.
Description of its algorithmic implementation, is beyond the scope of this paper and it will be
presented elsewhere~\cite{SP12}.

A validation of the method and demonstration of its capabilities
has been extensively studied in a number of toy models and in
two physical problems in hadronic physics both of current interest:
A) The extraction from lattice simulation
data of the masses of the spectrum of mesons and
baryons and B) The extraction of multipole amplitude strength from
nucleon electroexcitation spectra.  In all cases  studied the AMIAS
method recovered with the expected statical accuracy the input parameters
in the case of pseudodata. The derived uncertainties are compatible
with those obtained using the "jackknife"  technique~\cite{jackknife}
in the case of toy models and the lattice data.

We present below results with pseudodata to demonstrate the validity for the two cases mentioned above; analysis of physical data or simulations corresponding to these case have been published elsewhere.
\\

\textbf{A. Extraction of Mass Spectrum of Baryons from
Euclidean Time Correlators}

There has been impressive progress in lattice QCD calculations
where new algorithms and faster computers make feasible
high-precision simulations close to the physical parameters~\cite{Durr}.
As in the case of experimental data, simulation data are
characterized by statistical uncertainties and systematic error.
To extract physical quantities from lattice simulations such as
masses of hadrons, decay constants and form factors, fits to the
simulated data are performed. As in other fields, various
approaches have been explored in order to extract the physics of interest~\cite{MEM, Morningstar}.

We have successfully applied the AMIAS method to the analysis of
two-point correlators which result from calculations in
Lattice QCD~\cite{APS08}. Its generalization to more complicated objects such as
three-point correlators is under study.
Results extracted from lattice QCD simulations have been presented and compared
to traditional methods elsewhere~\cite{Lepage}.

In Lattice Gauge theories, the Euclidean time correlator $C(t)$ of
an interpolating operator $J({\bf x},t)$ and its spectral
decomposition for zero three-momentum is:
\begin{eqnarray}\label{eqn:correlator}
C(t)=\sum_{\bf x}<J({\bf x},t) J^\dagger({\bf 0},0)>=\sum_{n=0}^\infty A_n e^{-m_nt}
\end{eqnarray}
where the brackets denote the vacuum expectation value.
The exponential dependence is correct for Dirichlet boundary conditions.
In the large $t$ limit the state with  the lowest mass (ground state)
dominates the time dependence of the correlator. Fitting the asymptotic behavior of
$m_{\rm eff}(t)=-log\{C(t)/C(t+1)\}$ to a constant yields
the lower mass of the hadron while determination of higher masses gives the excitation energies of states
of the same quantum numbers as the ground state.

The case of lattice QCD simulations present an excellent case
for AMIAS.  As required, a framework that connects the data
and the model parameters of interest, the
masses ${m_j}$ and the overlap amplitudes ${C_j}$, is explicit and
in this case is  given by Eq.~\ref{eqn:correlator}.

We present here
a simple case employing pseudodata so as to demonstrate the validity and some features of the method.
Pseudodata were generated for the a system for a theory with
$f(t)= C_0 \exp(-m_0t) +  C_1 \exp(-m_1 t) $ and relative errors
that grow with time resembling lattice data. We have arbitrarily
chosen $C_0=1.0$, $m_0=0.500$, $C_1=3.0$ and $m_1=1.00$.
We have demonstrated that  the extracted values  have a precise meaning
through the analysis of pseudo-data generated with predetermined statistical accuracy.

\begin{figure}[h]
  \includegraphics[height=9.0cm, width=8.5cm]{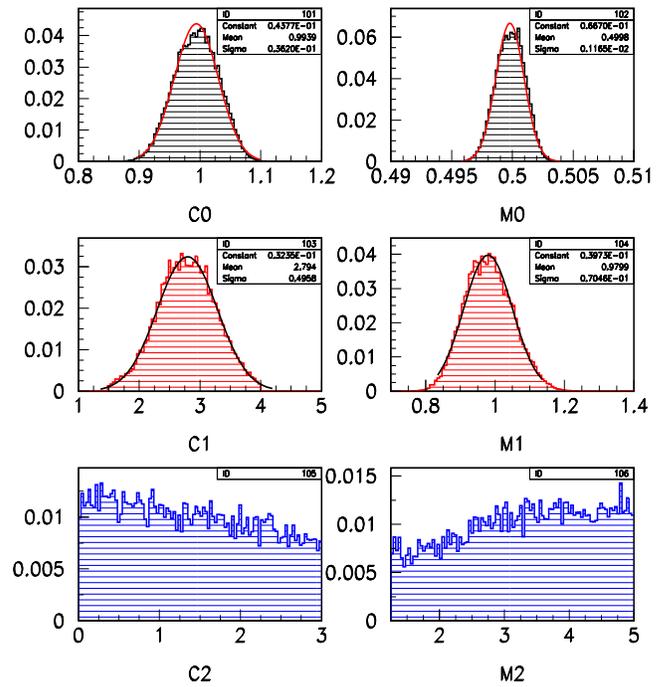}
  \caption{AMIAS generated Probability Distribution Functions (PDFs) for
  the masses of the ground state and excited states of the nucleon; The masses used to generate the pseudodata  are accurately reproduced. }
  \label{fig:Lattice_simulation}
\end{figure}

The extracted results are statistically compatible irrespective
of whether they were derived by taking $n=2$ or $n=3$ or $n > 3$.
The uncertainty of the fitted parameters grows as the number of the (a priory unknown) terms fitted is increased.
We adopted an ansatz whereby the number of terms employed is greater by one to those that can be extracted with finite uncertainty.
Similarly the size of the phase volume that the Monte-Carlo  method
is sampling does not affect the solution provided that the volume
is sufficiently large to include all "good solutions".
By "good solutions" we denote solutions with small or reasonable
$\chi^{2}/$(degrees of freedom).
As shown in Fig.~\ref{fig:Lattice_simulation} the parameters are accurately extracted,
in complete agreement with the generator values within the stated
statistical accuracy. As expected, search for $M2$ yields a null result.

AMIAS has been used to analyze lattice simulation data and the derived
results~\cite{APS08} that compare favorably to those derived by traditional methods
considered as defining the "state of the art" in the lattice community~\cite{Davis_Alexandrou}.
\\

\textbf{B. Extraction of Multipole Amplitudes from
electroexcitation Spectra}

The problem of extracting multipole amplitudes from
electroexcitation spectra with reduced model uncertainty
motivated the work that is reported here. In particular, the
verification of the conjecture that hadrons are non spherical~\cite{CNPAMB,RMP} has been demanding the
isolation with high precision of the small resonant quadrupole
amplitudes in the $N\rightarrow \Delta$ transition.
It was observed that increased accuracy in the experimental data would not yield more precise results~\cite{CNPAMB},
which were inherently limited by the limitations of the analysis methods employed.
This important case  provides a typical problem of high complexity, amenable to being solved by the AMIAS method.

The parameters of the model $\{A_{\nu}\}$, are the multipole amplitudes such as the $M^{1/2}_{L\pm}, M^{3/2}_{L\pm}$,
using standard spectroscopic notation. They relate to the data, cross sections and polarization observables in electroexcitation experiments,
through the CGLN formulation of the resonance electroexcitation~\cite{CGLN}.
They are infinite in number, so a truncation is needed and they are related to the data through a very complex convoluted scheme,
unlike the case of masses in QCD lattice data.
Furthermore, in this case, the parameters of the problem are subjected to the constraint of unitarization.

The validity of the AMIAS method was demonstrated with the employment of
pseudodata generated through the well established MAID scheme~\cite{MAID}
(which implements the CGLN formalism). AMIAS derived results have been demonstrated to
have a precise meaning through the analysis of pseudodata which
were generated with predetermined statistical accuracy. In the
cases presented below, we have frozen the $A_{\nu}^{1/2}$ and we
have varied the $A_{\nu}^{3/2}$ helicity amplitudes. Few demonstrative
cases of the pseudodata validation are presented below.

\begin{table}[tcb]
\caption{The multipole values extracted, in units of $10^{-3}/
m_{\pi}$, from two pseudodata sets are compared to the generator
(modified MAID) values. They are shown to be entirely compatible
with increasing precision in the extracted parameters as the
statistical accuracy of the data increases.}
\label{tab:pseudovalues}
\begin{tabular}{|c|r|r|r|r|}
  \hline
  Multipole &     Generator  & Set A $\quad\quad  $ & Set B $\quad\quad   $ \\
  \hline
  $M_{1+}$  & $27.248\ $ & $27.23\ \pm\ 0.13\ $ & $27.249\ \pm\ 0.001$ \\
  $L_{0+}$  & $ 3.500\ $ & $ 3.70\ \pm\ 0.23\ $ & $ 3.502\ \pm\ 0.002$ \\
  $L_{1+}$  & $ 1.048\ $ & $ 1.03\ \pm\ 0.08\ $ & $ 1.048\ \pm\ 0.001$ \\
  $E_{1+}$  & $ 1.481\ $ & $ 1.49\ \pm\ 0.18\ $ & $ 1.482\ \pm\ 0.002$ \\
  $E_{0+}$  & $ 4.225\ $ & $ 3.68\ \pm\ 1.02\ $ & $ 4.239\ \pm\ 0.013$ \\
  $M_{1-}$  & $ 4.119\ $ & $ 4.47\ \pm\ 1.31\ $ & $ 4.124\ \pm\ 0.013$ \\
  $L_{1-}$  & $ 1.205\ $ & $ 1.05\ \pm\ 0.43\ $ & $ 1.203\ \pm\ 0.008$ \\
  $E_{2-}$  & $ 1.024\ $ & $ 1.07\ \pm\ 0.45\ $ & $ 1.027\ \pm\ 0.006$ \\
  $L_{2+}$  & $ 0.007\ $ & $ 0.02\ \pm\ 0.01\ $ & $ 0.008\ \pm\ 0.001$ \\
  $E_{2+}$  & $ 0.006\ $ & $ 0.01\ \pm\ 0.01\ $ & $ 0.007\ \pm\ 0.001$ \\
  \hline
\end{tabular}
\end{table}

\begin{figure}[tcb]
  \includegraphics[height=6.0cm, width=8.1cm]{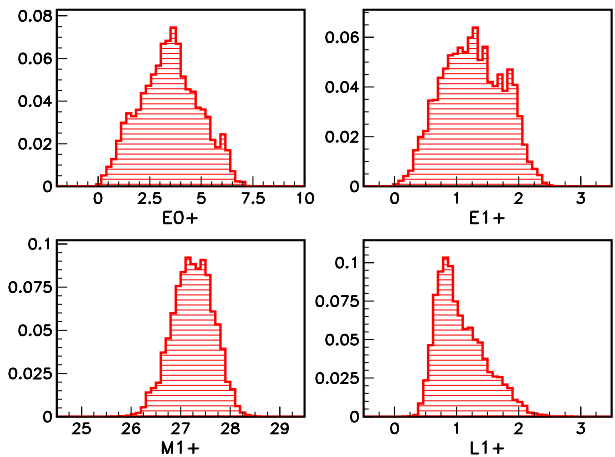}
  \caption{PDFs for the norms of some of the sensitive
  amplitudes of the analyzed Bates/Mainz experimental data set. The distributions allow
  the determination of the central value and corresponding uncertainty for
  each of the multipoles.}
  \label{fig:bates_sens_ampl}
\end{figure}

We use pseudodata with kinematics of the $Q^{2}=0.127$ (GeV/c)$^2$~
Bates and Mainz $N\rightarrow \Delta$ data~\cite{sp05} to demonstrate the validity
of the analysis.
The data set is published, is well understood and it is well described by MAID.
Two sets of pseudodata were generated using MAID, characterized by different
statistical accuracy: "Set A" with statistical accuracy similar
to that of the experimental values and "Set B"   with statistical
accuracy hundred times better than that of the experimental
values. These data were analyzed and the multipoles were
extracted which are tabulated and compared with the generator
values in the Table. We have tabulated only
extracted values which are derived with uncertainties better than
100\% for "Set A". It can be seen that the AMIAS extracted multipole values are
in complete agreement with the generator
values within the stated statistical accuracy. Also, as required,
the quoted uncertainties are reduced in set "B" (hundredfold),
proportionally to the statistical accuracy of the pseudodata sets.
For comparison, in Fig.~\ref{fig:bates_sens_ampl} the probability distributions is shown
for the most sensitive amplitudes of the Bates/Mainz experimental data set
analyzed with AMIAS.

To verify the ability of AMIAS to extract uncertainties which have precise statistical interpretation
is generally more difficult. The scaling behavior exhibited by the two sets of
pseudodata, discussed above, is a necessary but not sufficient condition.
The definitive validation was achieved by introducing an arbitrary uncertainty, a "generator uncertainty"
to the nominal generator multipole values. Multiple sets of data generated by randomized input
within the allowed uncertainties of the generator parameters are recovered by AMIAS.
This demonstration exercise was performed both for simple functional forms (e.g. polynomial functions)
and complicated cases such as this one (multipole amplitudes in a CGLN formalism),
results of which have been presented in~\cite{SP07}.
Furthermore, in the case of polynomial functions and lattice QCD two-point functions,
 derived jackknife errors are found to be statistically compatible with AMIAS uncertainties.

In summary: a novel method of analysis is shown to offer significant advantages
over existing methods in determining physical parameters from experimental or simulation
data: it is computationally robust, it provides methodolody independent answers with maximal precision in terms of the derived Probabilty Distribution Function for each parameter.

We thank  Prof. E.~Manousakis for enlightening discussions on the use of statistical theory
and  Prof. C.~Alexandrou for suggesting the use of AMIAS for lattice QCD gauge simulations.

\end{document}